\documentstyle{mn} 
\input psfig
\psfull
\def\kms{km\thinspace s$^{-1}$ }     
\def\degg{\ifmmode^\circ \else$^\circ $\fi } 
\def\et{{\it et~al.~}}
\newcommand{\ie}{{\it i.e.\ }}
\newcommand{\dg}{\nobreak^\circ }
\newcommand{\eg}{{\it e.g.\ }}
\newcommand{\simlt}{\mbox{$\stackrel{<}{_{\sim}}$} }
\newcommand{\simgt}{\mbox{$\stackrel{>}{_{\sim}}$} }
\newcommand{\lta}{{\small\raisebox{-0.6ex}{$\,\stackrel {\raisebox{-.2ex}{$\textstyle <$}}{\sim}\,$}}}
\newcommand{\gta}{{\small\raisebox{-0.6ex}{$\,\stackrel {\raisebox{-.2ex}{$\textstyle >$}}{\sim}\,$}}}
\title[10 GHz observations]
{10 GHz Tenerife CMB observations at $8\dg$ resolution and their
analysis using a new maximum entropy method}
\author[A.W. Jones et al.]
       { A.W.Jones$^1$, S. Hancock$^1$, A.N. Lasenby$^1$, R.D. Davies$^2$,
C.M. Guti\'errez$^3$, \and
G. Rocha$^1$, R.A. Watson$^{2,3}$ and R. Rebolo$^3$ \\ 
  $^{1}$Mullard Radio Astronomy Observatory, Cavendish Laboratory, 
        Madingley Road, Cambridge CB3 OHE, UK\\
  $^{2}$University of Manchester, Nuffield Radio Astronomy 
        Laboratories, Jodrell Bank, Macclesfield, Cheshire, SK11 9DL, UK\\
  $^{3}$Instituto de Astrof\'\i sica de Canarias, 38200 La Laguna,
        Tenerife, Spain 
  }
\date{}
\begin{document}
\maketitle
\begin{abstract}
The complete set of data from the Tenerife 10 GHz ($8\dg$ FWHM)
twin-horn, drift scan experiment is described. These data are
affected by both long-term atmospheric baseline drifts and short term
noise. A new maximum entropy procedure, utilising the time invariance
and spatial continuity of the astronomical signal, is used to achieve
a clean separation of these effects from the astronomical
signal, and to deconvolve the effects of the beam-switching.
We use a fully positive/negative algorithm to 
produce two-dimensional maps of the intrinsic sky fluctuations. 
Known discrete sources and
Galactic features are identified in the deconvolved map. The data from
the 10 GHz experiment, after baseline subtraction with MEM, is then
analysed using conventional techniques and new constraints on Galactic
emission are made.
\end{abstract}


\section{Introduction} 

Anisotropy of the cosmic microwave background (CMB) radiation provides one of
the key constraints against which cosmological models can be tested.
However, the detection of anisotropy at a level $\Delta T/T$ $\sim
10^{-5}$ is a challenging observational problem.  Attainment of these
sensitivity levels is only possible by making differential
measurements, switching between two different sky patches or by using
interferometric techniques.  The data
from the telescope consist of the true sky brightness distribution
convolved in some instrument beam, with an additional random noise
contribution. Combined with the often non-uniform scan strategies of
CMB experiments this makes the task of deconvolution to obtain a
two-dimensional map of the intrinsic fluctuations a non-trivial
proposition. Here we describe the analysis procedures that have been
developed and applied to the Tenerife drift scan data (Davies \et 
1987\nocite{davies87}, Watson \et 1992\nocite{watson92}, Hancock \et 
1994\nocite{h94}, Guti\'errez \et 1997) in order to
produce 2-D maps of the intrinsic fluctuations, while at the same time
providing scans free from atmospheric baseline variations. Although
the implementation described here is specific to the Tenerife
instruments, some of the techniques are more generally applicable to other CMB
data sets (see e.g. Maisinger {\em et al} 1997) and offer
a useful means of comparing between observations from
telescopes with different beam patterns and scan strategies.

In order to demonstrate the technique we consider here the analysis of
the total data set from the original (FWHM$\sim 8\dg$) Tenerife twin
horn radiometer experiment (Davies \et 1992\nocite{davies92}).
Although the instrument configuration differs from that of the current
telescopes (Davies \et 1996a\nocite{davies95}) and is less sensitive to
cosmological signals (Watson \et 1992\nocite{watson92}), this experiment
has surveyed a substantial fraction of the full sky, making it
interesting to attempt a 2-D mapping.
While a partial
analysis of a limited RA range, along a strip at Dec $= +40\dg$ has been
given elsewhere (Davies \et 1987\nocite{davies87}), we here present a
thorough analysis of all of the data, covering a selection of
declinations ranging from $-15\dg$ to $+45\dg$. At the operating frequency of
10.4 GHz our maps will be sensitive to synchrotron and free-free
emission in addition to the CMB and discrete radio sources. These maps
can be used to constrain contaminating signals in higher
frequency observations by virtue of the differing spectral dependencies
of the CMB and the foregrounds.  Testing the procedures on known
structures will also give us improved confidence when analysing the
higher frequency 15 and 33 GHz data to obtain 2-D maps of the CMB.

In Section 2 we present 
an analysis scheme utilising a maximum entropy based regularising function, 
to enable a clean separation of the astronomy
from spurious atmospheric effects and to provide a deconvolved 2-D
image of the microwave sky. In Section 3 we present details of the
observations and the implementation of the algorithm.
In Section 4 we test the positive/negative maximum entropy method with simulations of
the observations. Section 5 details the application of the algorithm
to the data from the Tenerife experiment. Sections 6, 7 and 8 describe the
separation of cosmological and astronomical signals,
the application of suitable significance tests and the interpretation
and conclusions that were reached.

\section{Maximum Entropy deconvolution}
\label{memalg}

In general the data from a CMB experiment will take the form of the true 
sky convolved with the instrumental response matrix with any baseline 
variations or noise terms added on. 
We assume that the observations obtained from a particular experiment 
have been integrated into discrete bins. For the $i$-th row
and the $j$-th column
the data, $y_i^{(j)}$, recorded by the instrument can be
expressed as the instrumental response matrix $R_i^{(j)}(i',j')$ acting
on the true sky $x(i',j')$:

\begin{equation}
\label{eq:data}
y_i^{(j)}= \sum_{(i',j')} R_i^{(j)}(i',j') x(i',j') +\epsilon_i^{(j)},
\end{equation}

\noindent
where $i'$ and $j'$ label the true sky row and column position respectively.
The $\epsilon_i^{(j)}$ term represents a noise term, assumed to be
random, uncorrelated Gaussian noise.  

It is immediately clear from Equation \ref{eq:data} that the inversion
of the data $y_i^{(j)}$ to obtain the two-dimensional sky distribution
$x(i',j')$, is singular. The inverse $R^{-1}$ of the instrumental
response function does not exist, unless the telescope samples all of the
modes on the sky and consequently there is a set of signals,
comprising the annihilator of $R$, which when convolved with $R$ gives
zero. Furthermore, the presence of the noise term $\epsilon$ will
effectively enlarge the annihilator of $R$ allowing small changes in
the data to produce large changes in the estimated sky signal. It is 
therefore necessary to use a technique that will approximate this inversion. 

\subsection{Positive and negative data reconstruction}

The method that we adopt is based upon the Maximum Entropy Method
(hereafter MEM) which is described in more detail in a companion paper
(Maisinger {\em et al} 1997). 
The problem with the MEM method in its
standard form is that it contains a logarithmic term that does not
allow the inclusion of negative features in the data. 
Laue, Skilling and Staunton (1985\nocite{laue85}) proposed 
a two channel MEM, which involved splitting the data into positive and
negative features and then reconstructing each separately but not taking into
account any continuity constraint between the two. This method
is inappropriate for differencing experiments as the positive and negative 
features originate from the beam shape and not from separate sources. 
White and Bunn (1995\nocite{whitemem}) have 
proposed adding a constant onto the data
to make it wholly positive. They use the Millimetre-wave Anisotropy 
Experiment (MAX) data to reconstruct a $5\dg \times 2\dg$ region of sky. 
As simulations we have performed show, this method introduces a bias 
towards positive (or negative if the data are inverted) 
reconstruction. The reason for this is that the added constant 
has to be small enough so that numerical errors are not introduced into
the calculations but a lower constant will give less range for the 
reconstruction of negative features and so the most probable sky will 
be a more positive one. We propose a new method to overcome both of these 
problems.

We consider the image to be the difference between two positive, additive 
distributions:
\[
x(\xi)=u(\xi)-v(\xi),
\]

\noindent
so that the expression for the cross entropy becomes:
 
\begin{equation}
S=\sum_{i',j'}  \left[\psi_{i',j'}-2m_{i',j'}-x_{i',j'} \ln \left( \frac{\psi_{i',j'}+ x_{i',j'}}{2m_{i',j'}} \right) \right],
\label{eq:entropy}
\end{equation}

\noindent
where $\psi_{i',j'}=(x_{i',j'}^2+4m_{i',j'}^2)^{1/2}$. The entropy
term constitutes our prior information about the fluctuations and
represents the minimum amount of information obtainable from the data.
$m_{i',j'}$ can be considered as a level of `damping' imposed 
on $x_{i',j'}$ rather than a default model as in standard MEM. 
A large value of $m$ allows large
noisy features to be reconstructed whereas a small value of $m$ will not
allow the final sky to deviate strongly from the zero mean.
However, $m$ does not include any correlations between pixels and so
no `smoothing scale' is introduced into the reconstruction. The final
reconstruction will contain no information on small angular scales
because the experiment is not sensitive to these scales and not
because the MEM approach biases the data. 
The detailed derivation of Equation \ref{eq:entropy} 
can be found in Maisinger \et (1997).
In a Bayesian sense we have now defined our prior and we can
calculate the probability of
obtaining our reconstructed sky $x$ (the hypothesis) given $y$ 
(the data):

\begin{equation}
\Pr(x | y) \propto \Pr(x) \Pr(y | x),
\end{equation}

\noindent
We can maximise $\Pr(x | y)$ to obtain the most likely 2-D image of the
microwave sky.

\section{Application to the data}
\label{obs}

As a demonstration of the power of maximum entropy 
we will now apply it to the beam switching experiment
sited at 2400m on the island of Tenerife. 
The old version of the instrument operated at 
10.4 GHz during the years 1984-85 and 
is a drift scan experiment (Davies \et 1992). 
Repeated scans were built up at a set of
declinations, $-17.3$, $-2.4$, $+1.1$, $+7.3$, $+17.5$, $+27.2$, $+37.2$, $+39.4$,
$+42.6$ and $+46.6$ degrees, with the deepest integrations centred on Decs
$+1.1\dg$ and $+39.4\dg$.

Approximately half of the time was devoted to maintenance and
calibration runs, while further observing time was lost due to poor
weather conditions.  The instrument was left running over a continuous
period of up to $2 \times 24$ hours and thus a single scan contains a
maximum of 2 full coverages in right ascension, with data being taken
over a period of up to 3 calendar days.  Data were taken by adjusting
the wagging mirror to observe the chosen declination and allowing the
Earth's rotation to sweep the beams in right ascension. Each wagging
cycle consisted of a 4 second integration with the beams directed to
the East, followed by a 4 second integration to the West. At each
position the beam difference, $(T_C - T_E) \pm \sigma_E$ or $(T_W - T_C)
\pm \sigma_W$ (C, E and W denote the centre, East and West beam
positions respectively), and then the corresponding
difference in beam difference 

\begin{equation} 
(T_C - \frac{1}{2} (T_E + T_W)) \pm \sigma 
\label{eq:secdiff} 
\end{equation} 

\noindent
where $\left(\sigma={1\over 2}\sqrt{\sigma_E^2+\sigma_W^2} \right)$,
were calculated. Over each 82 second cycle, containing 6 pairs of
secondary differences, the second difference and its standard deviation
were recorded, along with a calibration signal.  The data were
calibrated and edited as described in Davies \et (1992), resulting in a final
data set with the properties given in Table \ref{ta:8deg} which shows
the declinations surveyed, the number of coverages at each declination
and the mean and standard deviation of the scan lengths in hours. The
data for each scan were binned in $1\dg$ intervals in RA to convert
them to a more tractable form. Binning the scans reduces
the effects of short term receiver and atmospheric noise, but does not
affect the long term drifts seen in individual scans.

\begin{table}
\caption{Observations with the $8.3^{\circ}$ FWHM 10.4 GHz experiment.}
\label{ta:8deg}
\begin{center}
\begin{tabular}{|c|c|c|c|} \hline
Declination & Number & Mean scan & RMS length \\ 
 & of scans & length (hours) & (hours) \\ 
$+46.6^{\circ}$ & 15 & 14.1 & 3.1 \\
$+42.6^{\circ}$ & 16 & 22.0 & 9.7 \\
$+39.4^{\circ}$ & 42 & 14.1 & 6.3 \\
$+37.2^{\circ}$ & 18 & 14.6 & 7.6 \\
$+27.2^{\circ}$ & 17 & 11.3 & 3.9 \\
$+17.5^{\circ}$ & 16 & 21.1 & 10.3 \\
$+07.3^{\circ}$ & 13 & 9.7 & 3.3 \\
$+01.1^{\circ}$ & 52 & 15.8 & 11.1 \\
$-02.4^{\circ}$ & 6 & 10.7 & 4.9 \\
$-17.3^{\circ}$ & 20 & 8.4 & 4.2 \\
\hline
\end{tabular}
\end{center}
\end{table}

\subsection{The data scans: long period baseline drifts.}
\label{longbase}

\begin{figure}
\centerline{\hbox{\psfig{figure=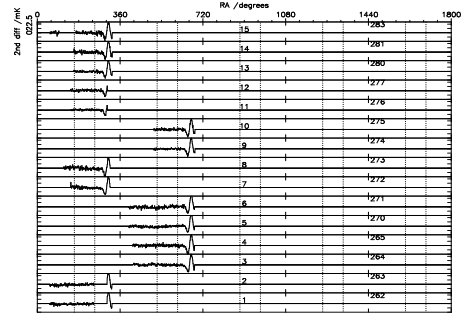,height=2.0in,width=3.0in,angle=-90}}}
\caption{The 15 scans obtained at Dec $=46.6\dg$ 
displayed as a function of right ascension. Each plot shows 
the second difference in mK after binning into $1\dg$ 
bins. A running mean has been subtracted from each scan. 
Long scans are displayed modulo $360\dg$.}
\label{fig:typical15}
\end{figure}

\begin{figure}
\centerline{\hbox{\psfig{figure=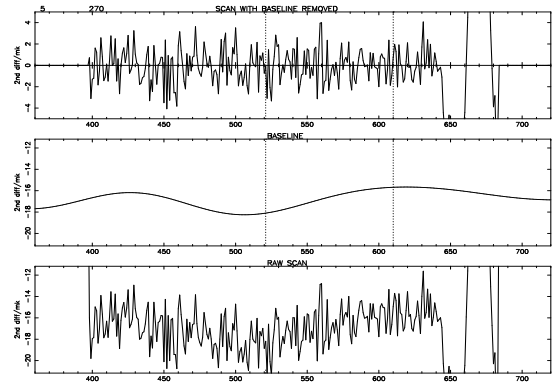,height=2.0in,width=3.0in,angle=-90}}}
\caption{The data from scan 5 of Figure 1.
displayed on an expanded temperature scale against RA bin number. Long 
timescale variations in the mean level are evident in the RAW scan
(bottom panel). The middle panel shows the 
baseline fit found by the method of Section 2.
The top panel shows the
baseline corrected scan. The bin numbers exceed 360 since 
the scan begins near the end of an LST day, and the data are not folded modulo 
$360\dg$.}
\label{fig:scan5}
\end{figure}

As an illustration of the nature of the data, we show in
Figure~\ref{fig:typical15} the set of all scans for Dec $=+46.6\dg$.
Immediately evident at RA $\approx 308\dg$ is the strong Galactic plane
crossing with characteristic shape due to the triple beam pattern,
while at RA $\approx 70\dg$ a weaker crossing can just be discerned. A
constant term has been subtracted from each scan in order to bring the
non-Galactic plane sections to a mean of approximately zero. However,
in \eg scan 5 of this set (numbered upwards from the bottom), which is
shown on an increased scale in Fig.~\ref{fig:scan5}, a slow variation
in baselevel, amplitude (peak to peak) $\sim 2$ mK, is distinctly
evident. As discussed in Davies \et (1992), most of the scans obtained show these
variations, to a greater or lesser degree, and therefore their removal
is an important part of the analysis. These long period baselines vary
both along a given scan and from day to day, clearly indicating that
they are due in the main to atmospheric effects, with a possible
contribution from diurnal variations in the ambient conditions. We note
that the timescale for these baselevel variations appears to be several
hours.  We examine this quantitatively by calculating the transfer
function of the experiment, which defines the scales of real structures
on the sky to which the telescope is sensitive. Variations produced on
scales other than these will be entirely the result of non-astronomical
(principally atmospheric) processes and should be removed.

As the Earth rotates the beams are swept in RA over a band of sky
centred at a constant declination.  For our present illustrative
purposes, it is sufficient to approximate the beams as one-dimensional
in RA, with the beam centre and the East and West throw positions lying
at the same declination.  The beamshape for each individual horn is
well represented by a Gaussian with dispersion $\sigma=$FWHM$/2\sqrt{2\ln 2}=3.57\dg$:

\begin{equation}
B(\theta) = \exp \left(- \frac{\theta^2}{2 \sigma^2} \right),
\end{equation}

\noindent
and the beam switching in right ascension, 
may be expressed as a combination of delta
functions:

\begin{equation} \label{eq:switch} S(\theta)= \delta(0)- \frac{1}{2}
(\delta(\theta_b) + \delta(- \theta_b)), 
\end{equation}

\noindent
for a switch angle $\theta_b=8.3\dg$ in RA. So, the beam pattern is
\[P(\theta)=B(\theta)*S(\theta). \] Thus, the transfer function, (\ie
the Fourier transform of the beam pattern) is just:

\begin{equation}
g(k)=2 \sqrt{2 \pi}\sigma \exp \left( \frac{- k^2 \sigma^2}{2} \right) \sin^2 \left( \frac{k \theta_b}{2} \right) .
\end{equation}

\begin{figure}
\centerline{\hbox{\psfig{figure=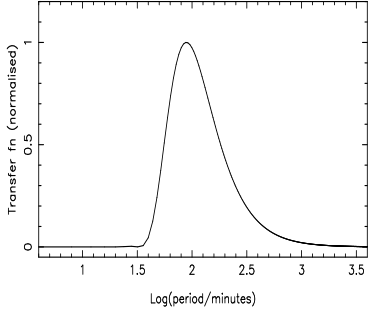,height=2.0in,width=2.0in,angle=-90}}}
\caption{The transfer function for the Tenerife experiments.}
\label{fig:tranfn}
\end{figure}

In Figure \ref{fig:tranfn}, the transfer function for waves of period
$\theta= 2 \pi/k$ is plotted. As a function of declination the $\theta$
co-ordinate must be multiplied by a factor of $\sec \delta$ because a
true angle $\theta$ on the sky covers $\sim \theta / \cos \delta$ in right
ascension.  The peak response of the instrument is to plane waves of
period $\sim 22\dg \sec \delta$, \ie individual peaks/troughs with FWHM
$\sim 7\dg$.  The response drops by a factor 10 for structures with
periods greater than $\sim 7$ hours and less than $\sim 40$ minutes.
The long period cutoff is due to cancellation of the large-scale
structures in the beam differencing pattern, while the short period
cutoff is simply due to dilution of structures within a single beam.
The cutoff on large scales in particular is significant for the
analysis, since it tells us that variations in the data on timescales
$\simgt 7^{h}\sec\delta$ are almost certainly due to long timescale
atmospheric effects, or terrestrial and environmental effects, rather
than being intrinsic to the astronomical sky. 
Thus identification and
removal of such `baseline' effects (Davies \et 1992) is important. 
By using the whole data set to calculate
the most probable astronomical sky signal with maximum entropy deconvolution,
we can simultaneously fit a long-timescale Fourier component baseline to 
each scan. We can then stack together 
$n$ days of data at a given declination to obtain a final scan with a
$\sim \sqrt{n}$ improvement in sensitivity to true astronomical features.

\subsection{The beam}
\label{tbeam}

Note that in the
case of the Tenerife experiments, it is not
necessary to re-define the beam matrix for each position in RA (which
requires a matrix $R_i^{(j)}$ for the $i$-th bin in RA and $j$-th bin
in declination) since the beam is translationally
invariant in the RA direction. However, due to the large sky coverage
and spherical nature of the sky it is clear that the beam shape
projected into RA and Dec co-ordinates will be a function of
declination. We can write the beam matrix, $R^{(j)}(i',j')$, 
for declination $j$ as follows,
 
\begin{equation}
R^{(j)}(i',j')=
\label{eq:beam}
\end{equation}
\nopagebreak
\begin{displaymath}
C \left[ \exp \left(-\frac{\theta_C^2}{2 \sigma^2} \right )-
\frac{1}{2} \left( \exp \left(-\frac{\theta_E^2}{2 \sigma^2} \right )
+ \exp \left(-\frac{\theta_W^2}{2 \sigma^2} \right ) \right )  \right ]
\end{displaymath}

\noindent 
where $\theta_C$, $\theta_E$, and $\theta_W$ represent the true angular
separation of the point $(i',j')$ from the beam centre and the East and
West throw positions respectively. The normalisation of the beam matrix
is determined by $C$ and 
is implemented with respect to a single beam.
The angles  $\theta_C$, $\theta_E$, and $\theta_W$ can be calculated
using spherical geometry. If the beam is centred at
Dec. $\delta^{(j)}$ and we let
$\alpha_0$ be the (arbitrary) RA origin for the definition of all the
beams, for a source at a general $(\alpha,\delta)$ corresponding to the
grid point $(i',j')$ we have a distance from the main beam centre of

\begin{equation}
\theta_{C} = \arccos \left( \sin \delta^{(j)} \sin \delta + \cos (\alpha_{0} -
\alpha) \cos \delta^{(j)} \cos \delta \right).
\label{eq:thetac}
\end{equation}
 
There are also two other beams (with half the amplitude of the central 
beam), due to
the beamswitching and mirror
wagging, a distance $\theta_b$ (the beamthrow) either side of the
central peak. These have RA centres given by the $\alpha_{E}$ or
$\alpha_{W}$ in \[ \cos ( \alpha_{E {\rm \, or \, } W} - \alpha_{0}) =
\frac{\cos \theta_{b} - \sin^{2} \delta ^{(j)}} { \cos^{2}
\delta^{(j)}}, \] and (fairly accurately for the beamthrow used in
practice) declinations of $\delta^{(j)}$ still. Thus their angular distances
from the source at $(i',j')$ can be worked out from Equation \ref{eq:thetac}, yielding
$\theta_{E {\rm \, or \, } W}$ say, and the final $R^{(j)}$ entry
computed from Equation \ref{eq:beam}.

\subsection{Implementation of MEM}

For our data set tests have shown that we have random Gaussian noise, for
which the likelihood is given by:

\begin{equation}
\label{eq:chilike}
\Pr(y | x) \propto \exp \left( \frac{- \chi^2}{2} \right),
\end{equation}

\noindent
where $\chi^2$ is the misfit statistic (Equation \ref{eq:chisqu}).
Thus in order to maximise $\Pr(x | y)$ we simply need to minimise the
function

\begin{equation}
\label{eq:f}
F=\chi^2 - \alpha S
\end{equation}

\noindent
where we have absorbed a factor of two into the Lagrangian multiplier, $\alpha$.
Thus the process is to iterate to a minimum in $F$ by consistently
updating the 2-D reconstructed sky $x(i',j')$, while at the same time
building up a set of baselines for each scan. The baselines are
represented by a Fourier series

\begin{equation}
\label{eq:baseline}
b_{ir}^{(j)}=C_{0r}^{(j)} + \sum_{n=1}^{nmax} \left[ C_{nr}^{(j)} \cos(\frac{2n \pi i}{l}) + D_{nr}^{(j)} \sin(\frac{2n \pi i}{l}) \right] 
\end{equation}

\noindent
for the $r$-th scan at the $j$-th declination with $nmax$ baseline
coefficients to be fitted. The basis
data vector index $i$ runs from $1$ to 
$3 \times 24^{h}$. Thus to obtain a minimum period solution \simgt
${7^{h}} \sec \delta$ (Section \ref{longbase}) 
we must limit the number of baseline coefficients, $nmax$ in 
Equation \ref{eq:baseline} to less than 9 (for the case $\delta=40\degg$).

For the $r$th scan, Equation \ref{eq:data} may be written

\begin{equation}
y_{ir}^{(j)}=y_i^{pred(j)}+b_{ir}^{(j)}+\epsilon_{ir}^{(j)},
\end{equation}

\noindent
where the baseline variation $b$ has now been included and 

\begin{equation}
\label{eq:pred}
y_i^{pred(j)}= \sum_{i',j'} R^{(j)}(i',j')x(i'+i-i_0',j')
\end{equation}

\noindent
is the signal we predict to be produced by the telescope in the
absence of noise and baseline offsets. The beam matrix $R$ is now defined
with respect to an origin $i'_o$ in the $i'$ direction as it is
translationally invariant in RA. So, the $\chi^2$ for the problem is

\begin{equation}
\label{eq:chisqu}
\chi^2=\sum_{j=1}^{ndecs}\sum_{i=1}^{nra} w_{i}^{(j)} (y_i^{pred(j)}-y_{i}^{obs(j)})^2,
\end{equation}

\noindent
for a total number of declinations $ndecs$,
total number of RA bins $nra$ and observed data value $y_{i}^{obs(j)}$
with weighting factor $w_{i}^{(j)}$, which are a weighted average over the
$ns$ scans with the baseline $b_{ir}^{(j)}$
subtracted from each of the scans $y_{ir}^{(j)}$ ($r$ is an index running over the $ns$ scans).
In the absence of data
$w_{ir}^{(j)}$, for each individual scan, is set to zero and when data is present it is given by
the inverse of the variance for the data point.
It is possible to compute $y_i^{pred(j)}$, since we can use our knowledge of
the geometry of the instrument to calculate the expected response
function $R^{(j)}(i',j')$ for each $i'$, $j'$, at RA $i$ and declination $j$, thus
$\chi^2$ is fully defined.

If we know the value of the regularising parameter $\alpha$ and the
`damping' term $m$ then
we know $F$ and our best sky reconstruction is that for which $\partial
F /\partial x_{ij}=0$, $\forall x_{ij}$.  This is most easily
implemented by applying one-dimensional Newton-Raphson iteration
simultaneously to each of the $x_{ij}$ to find the zero of the function
$G(x)=\partial F  / \partial x$. This means that we update $x$ from the
$n$-th to the $(n+1)$-th iteration by

\begin{equation}
x_{lm}^{n+1} = x_{lm}^{n} -\gamma \left( \frac{G(x_{lm}^n)}{\left . \frac{\partial G}{\partial x_{lm}} \right |_{x_{lm}^n} } \right ).
\end{equation}

Convergence towards a global minimum is ensured by setting a suitable
value for the loop gain $\gamma$ and updating $x_{lm}$ only if $\left .
\frac{\partial G}{\partial x_{lm}} \right |_{x_{lm}^n}$ is positive (so
that progress is always towards a minimum).
By simultaneously fitting for the parameters of the baselines, it is
possible to calculate the best reconstruction of the microwave sky
along with an atmospheric baseline for each scan. To fit for the
baseline parameters $C_{0r}^{(j)}$, $C_{nr}^{(j)}$ and $D_{nr}^{(j)}$ as
expressed in Equation \ref{eq:baseline} it is sufficient to implement a
simultaneous but independent
$\chi^2$ minimisation on each of these to obtain the baseline for the
$r$-th scan. From the Bayesian viewpoint minimising $\chi^2$ is just
finding the maximum posterior probability by using a uniform prior. This is 
also done with a Newton-Raphson iterative technique with a new loop gain,
$\gamma_b$.

\subsection{Choosing alpha and m}

In this MEM approach, the entropic regularising parameter
$\alpha$ controls the competition between the requirement for a
smoothly varying sky and the noisy sky imposed by our data. The larger
the value of $\alpha$ the more the data are ignored.  The smaller the value of
$\alpha$ the more structure is reconstructed.  We wish to make a choice of 
$\alpha$ that will take maximum
notice of the data vectors containing information on the true sky
distribution, while using the `damping' constraint and the beam
sensitivity to reject the noisy
data vectors. In some sense, one may think of the entropy term as using
our prior information that the sky does not contain large
fluctuations at some level to fill in
for the information not sampled by the response function, thereby
allowing the inversion process to be implemented.

The optimum choice of $\alpha$ is somewhat controversial in the
Bayesian community and while several methods exist (Gull
1989\nocite{gull89}, Skilling 1989\nocite{skilling89}) it is difficult
to select one above the others that is superior. We use the 
criterion that $\chi^2-\alpha S=N$, where $N$ is the number of data points 
that we are trying to fit in the convolved sky. If any of the data 
points are weighted to zero, as the galactic plane crossing is in our
case, these points should not be included in $N$. 
Increasing/decreasing
$\alpha$ by a factor of ten decreases/increases the amplitude of the
fluctuations derived in the final analysis by \simlt 5 \%.  We decrease $\alpha$
in stages until $\chi^2-\alpha S=N$; experience has
shown that a convergent solution is best obtained with the typical
parameter values given in Table \ref{ta:params} for the data set considered
here. Below this value for 
$\alpha$ the noisy features in the data have a large effect and the 
scans are poorly fitted. Note that one cannot 
attach any significance to the absolute value of
$\alpha$, since it is a parameter that depends on the scaling of the
problem.

\begin{table}
\caption{The parameters used in the MEM inversion.}
\label{ta:params}
\begin{center}
\begin{tabular}{|c|c|} \hline
Parameter & Value\\ 
$\alpha$ & $2\times 10^{-2}$\\
$m$ & $10\mu$K\\
$\gamma$ & $0.01$\\
$\gamma_b$ & $0.05$\\
\hline
\end{tabular}
\end{center}
\end{table}

There is less constraint on the choice of $m$, the `damping' term. We choose
$m$ to be of similar size to the {\em rms} of the fluctuations so that the
algorithm has enough freedom to reconstruct the expected features. 
Increasing/decreasing $m$ by an a few orders of magnitude 
from this value does not alter the final 
result significantly so that the absolute value of $m$ is not important.
This is different to a positive only MEM because in that 
case $m$ is chosen to be the default model (the value of the sky 
reconstruction in the absence of data) and is therefore more 
constrained by the problem itself. In our case as $m$ is the default 
model on the two channels and not on the final sky there is a greater
freedom in its choice. 

\section{Testing the algorithm}
\label{sim}

Before applying the MEM algorithm to the real data, simulations were carried
out to test its performance. Two-dimensional sky maps were simulated using a
standard cold, dark matter model (${H_o}=50$\kms, $\Omega_b=0.1$)
with an {\em rms} signal of $22\mu$K per pixel
(normalised to COBE second year data, ${Q_{rms-PS}}=20.3\mu$K, see 
Tegmark \& Bunn 1995). Observations from the sky maps were
then simulated by convolving them with the Tenerife beam. Before noise was
added the positive/negative algorithm was tested by analysing the data 
and then changing the sign of the data and reanalysing
again. In both cases the same, but inverted, reconstruction was found for the MEM
output and so we conclude that our method of two positive channels introduces
no biases towards being positive or negative. Various 
noise levels were then added to 
the scans before reconstruction with MEM. 
The two noise levels considered here
are $100\mu$K and $25\mu$K on the data scans,
which represent the two extrema of the 
data that we expect from the various Tenerife configurations
($100\mu$K for the 10 GHz, FWHM=$8\dg$ data and $25\mu$K for the 15 and
33 GHz, FWHM=$5\dg$ data). 

\begin{figure}
\centerline{\hbox{\psfig{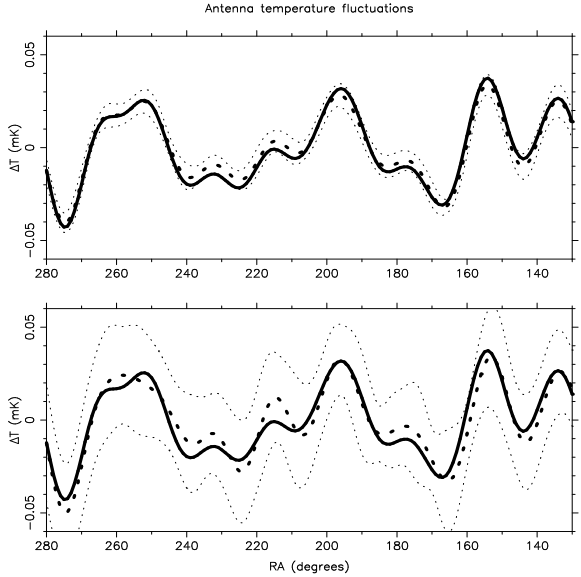}}}
\caption{The  solid line shows the sky simulation convolved
with the Tenerife
$8\dg$ experiment. The bold dotted line in the top figure 
shows the MEM reconstructed sky after
reconvolution with the Tenerife beam, averaged over 
simulations of the  MEM output from a simulated experiment with $25\mu$K
Gaussian noise added to each scan. Also shown are the $68\%$ confidence limits 
(simulation to simulation variation; dotted lines) on this
reconvolution. The bold dotted line in the bottom figure 
shows the reconvolution averaged over 
simulations with $100\mu$K Gaussian noise added to each scan. 
The $68\%$ confidence
bounds (dotted lines) are also shown for this scan.}
\label{fig:simscan}
\end{figure}

\begin{figure}
\centerline{\hbox{\psfig{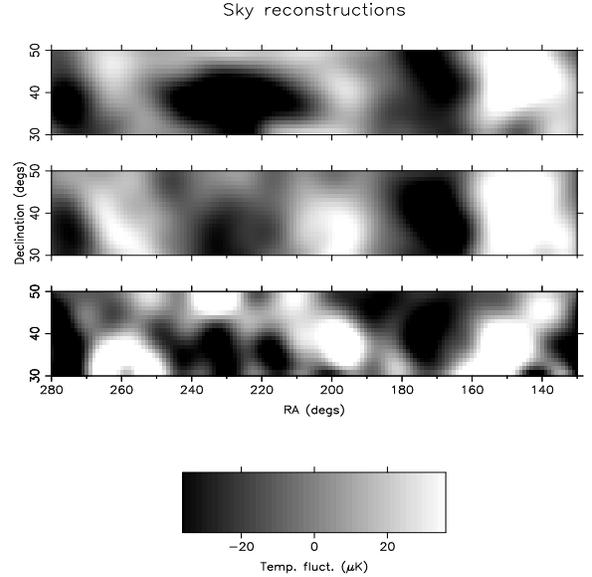}}}
\caption{The top figure is the simulated sky convolved with an $8\degg$
Gaussian beam. The middle and bottom figures are the reconstructed skies from
that simulation after scans with $25\mu$K and $100\mu$K noise levels
respectively, simulating the Tenerife experiment, were made and passed through
MEM. They are all convolved with a final Gaussian of the same size.}
\label{fig:3plot}
\end{figure}

Figure \ref{fig:simscan} shows the convolution of one simulation with the 
Tenerife beam and the result obtained from MEM with the two 
noise levels. The plots are averaged over 30 simulations and the bounds
are the $68\%$ confidence limits (simulation to simulation variation). As seen 
MEM recovers the underlying sky simulation to good accuracy for both 
noise levels, with the $25\mu$K result the better of the two as 
expected. Figure \ref{fig:3plot} 
shows the reconstructed intrinsic sky from two of the simulations after 60
iterations of the MEM algorithm  as compared with the real sky simulations 
convolved in an $8\degg$ Gaussian beam. Various common features are seen in 
the three scans like the maxima at ~RA $150\dg$, minima at ~RA $170\dg$ 
and the partial ring feature between RA $200\dg$ and $260\dg$ with 
central minima at ~RA $230\dg$, ~Dec. $+35\dg$. 
All features larger than the {\em rms} are reconstructed in both 
the $25\mu$K and $100\mu$K noise simulations. However, there is a some
freedom in the algorithm to move these features within a beam
width. This can cause spurious features to appear at the edge of the
map when the guard region (about $5\dg$) around the map contains a
peak (this can be seen in the map as a decaying tail away from the edge). 
For example, the feature at RA $230\dg$, Dec $50\dg$ has been
moved down by a few degrees out of the guard region 
in the $100\mu$K noise simulation so it
appears more prominently on the ring feature. 

There is a tendency for 
the MEM algorithm to produce superresolution (Narayan \& Nityananda 1986)
of the features in the sky 
so that even though the experiment may not be sensitive to small angular
scales the final reconstruction appears to have these features in it. Care
must be taken not to interpret these features as actual sky features but 
instead the maps should be convolved back down with a Gaussian to the 
size of the features that are detectable. This has been done with the 
two lower plots in Figure \ref{fig:3plot},
so that a direct comparison between all three is 
possible.  
By comparison of these plots we are confident in saying that the 
reconstructed sky obtained from the MEM algorithm does give us a good 
description of the actual sky. 

\begin{figure}
\centerline{\hbox{\psfig{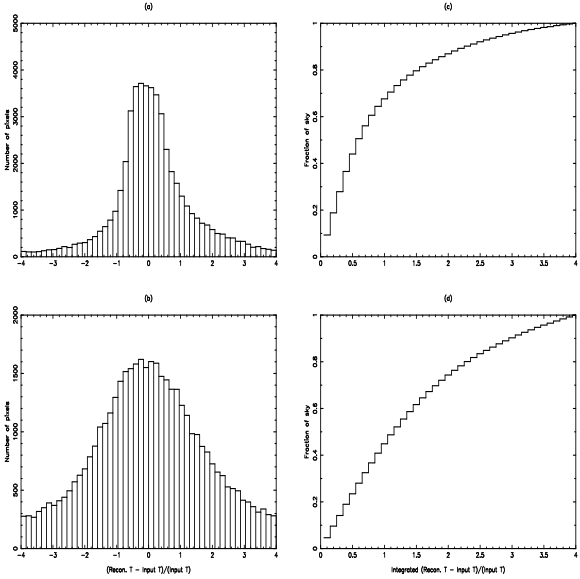}}}
\caption{ Histograms of the errors in the reconstruction of the
simulated sky maps. (a) shows the 
${{\rm T}_{recon}-{{\rm T}_{input}}}\over {{\rm T}_{input}}$ for the
$25\mu$K noise simulation and (c) shows the integrated 
$\left| {{{\rm T}_{recon}-{{\rm T}_{input}}}\over {{\rm T}_{input}} } \right|$ 
for (a). (b) and (d) are the corresponding plots for the
$100\mu$K noise simulation. }
\label{fig:histo}
\end{figure}

As an indicator of the error on the final sky reconstruction from the
MEM, a histogram of the fractional difference between the input and output map
temperatures is plotted in Figure \ref{fig:histo}. If the initial
temperature at pixel $(i,j)$ is given by ${\rm T}_{input}$ and the
temperature at the same pixel in the output reconstructed map (after
convolution with a Gaussian beam to avoid superresolution) is given
by ${\rm T}_{recon}$ then the value of

\begin{equation}
{{{\rm T}_{recon}}-{{\rm T}_{input}}}\over{{\rm T}_{input}}
\end{equation}

\noindent
is put into discrete bins and summed over all $(i,j)$. The final
histogram is the number of pixels within each bin. The output map
has been averaged over pixels within the beam FWHM as features can
move by this amount. A graph centred on -1 would mean that the
amplitude of the output signal is near zero while a graph centred on 0
would mean the reconstruction is very accurate. As can be seen both
graphs (Figure \ref{fig:histo} (a) and (c) for the $25\mu$K and
$100\mu$K noise simulations respectively) can be well
approximated by a Gaussian centred on a value just below zero. This
means that the MEM has a tendency to reconstruct the data with
slightly smaller amplitude which increases with the level of noise ($\sim
10\%$ smaller for the $25\mu$K simulation and $\sim 20\%$ for the $100\mu$K
simulation). This is expected from the entropy `damping' property.
From the integrated plots (Figure \ref{fig:histo} (b) and
(d)) we can expect to reconstruct all features with better than $50\%$
accuracy a half of the time for the $25\mu$K noise simulation and
a third of the time for the $100\mu$K noise simulation. The next section 
describes the implementation
of the algorithm for the actual $8\degg$ Tenerife experiment data.

\section{Reconstructing the sky at 10.4 GHz}
\label{recon}

To apply the deconvolution process described in Section \ref{memalg} to
the data in Section \ref{obs} one must decide on the required dynamic
range for the reconstruction and also select parameters that not only
achieve convergence of the iterative scheme, but also make the fullest
use of the data.  The amplitude of the fluctuations that we are
interested in is at least an order of magnitude smaller than the
magnitude of the signal produced during the major passage through the
Galactic plane region ($\sim 45$ mK at $\sim$ Dec. $+40\degg$). Clearly, any
baseline fitting and reconstruction will be dominated by this feature
at the expense of introducing spurious features into the regions in
which we are interested. For this reason the data
(Table~\ref{ta:nogal}) corresponding to the principal Galactic plane
crossing are not used in the reconstruction.

\begin{table}
\caption{The Galactic plane regions excised at each declination.}
\label{ta:nogal}
\begin{center}
\begin{tabular}{|c|c|} \hline
Declination & RA range excised (degrees)\\ 
$+46.6^{\circ}$ & 275-340\\
$+42.6^{\circ}$ & 275-340\\
$+39.4^{\circ}$ & 275-340\\
$+37.2^{\circ}$ & 275-340\\
$+27.2^{\circ}$ & 280-320\\
$+17.5^{\circ}$ & 265-310\\
$+07.3^{\circ}$ & 255-310\\
$+01.1^{\circ}$ & 255-310\\
$-02.4^{\circ}$ & 260-310\\
$-17.3^{\circ}$ & 255-300\\
\hline
\end{tabular}
\end{center}
\end{table}

In contrast, the anti-centre crossing ($\sim$RA $60^{\circ}$ at
$\sim$Dec. $+40\degg$ ) corresponding to scanning through the Galactic plane, but looking
out of the Galaxy, is at an acceptable level ($\lta 5$ mK) and is a useful check on the performance and consistency
of the observations.
With the parameters set as in Table \ref{ta:params}, 
$\chi^2$ demonstrates a rapid convergence. For example, the
change in $\chi^2$ after 120 iterations of MEM is
$\Delta \chi^2/\chi^2 \simeq -9\times 10^{-4}$ while the change in $\chi_{base}^2$ is
$\Delta \chi_{base}^2/\chi_{base}^2 \simeq -2\times 10^{-4}$. 
 
\begin{figure}
\centerline{\hbox{\psfig{figure=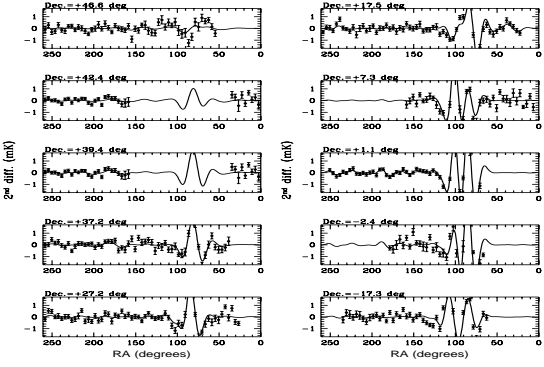,height=2.0in,width=3.0in,angle=0}}}
\caption{The stacked scans at each declination displayed as a function 
of right ascension. Again the plots are the second difference binned 
into $4\dg$ bins and the $68\%$ confidence limits. The main Galactic
plane crossing has been excluded, and only positions on the sky in
which we have more than $\sim$10 independent measurements have been
plotted. Also shown (solid line) is the reconvolved result from MEM overlayed 
onto each declination scan.}
\label{fig:stack}
\end{figure}

The fitted baselines are subtracted from the raw data set to
provide data free from baseline effects, allowing the scans for a given
declination to be stacked together to provide a single high sensitivity
scan.  However, problems arise when the baseline variations in the raw
data are so extreme as to prevent their successful removal in the MEM
deconvolution analysis. As noted in Davies \et (1996a), this problem
is exaggerated at the higher frequencies where the water vapour
emission is higher. At these higher frequencies
it is clear that the variations in baseline are, in certain cases, too 
extreme for removal 
and will therefore result in artefacts in the
final stacked scan. We are confident that these artefacts result from
poor observing conditions rather than being intrinsic to the astronomy,
because such problems occur only for days with severe baselines and
appear in a randomly distributed fashion for different days.  Removal
of such data is essential if one is to obtain the necessary sensitivity
to detect CMB fluctuations. This involves the laborious task of
examining each raw scan and its baseline and deciding if the data are
usable. In such cases where the data is un-salvageable, then the data
for the full $360^{\circ}$ observation are discarded. This ensures that
there is no bias introduced by selectively removing features in the
scans. After this final stage of editing, the baseline fitting must be
repeated for the full remaining data set. The MEM process will now be
able to search for a more accurate solution and will produce a new set
of more accurate baselines.
The coverages of a given declination can now be stacked together.
Figure \ref{fig:stack} shows the stacked results at each declination
compared with the reconvolution of the MEM result with the beam. The main Galactic 
crossing has been excluded from this data but the weak Galactic 
crossing is clearly visible at RA=50\degg-100\degg. Only data points on
the sky with more than ten independent measurements have been plotted
and in the absence of data the continuity between declinations has
clearly been
used by the MEM to reconstruct the scans. At lower declinations 
this crossing shows a complex structure with peak amplitudes $\sim$ a 
few mK. The data with better sensitivities are those at Dec.=+39.4\degg 
and +1.1\degg.

\begin{figure}
\centerline{\hbox{\psfig{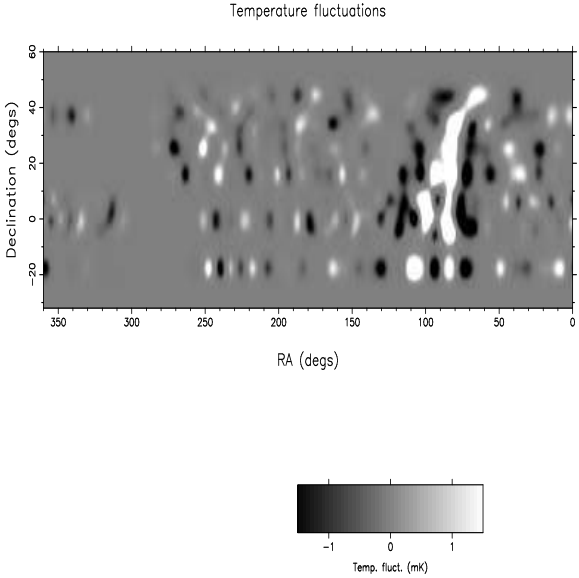}}}
\caption{MEM reconstruction of the sky at 10.4 GHz, as seen by the Tenerife
$8.4\dg $ FWHM experiment.}
\label{fig:8degpn}
\end{figure}

The sky is not fully sampled with this data set but
the MEM uses the continuity and `damping' constraints on the data to 
reconstruct a two-dimensional sky model. 
In Figure \ref{fig:8degpn}, the sky reconstruction is shown. Although
a rectangular projection has been used for display, the underlying
computations use the full spherical geometry for the beams (as
described in Section \ref{tbeam}).
The anti-centre crossings of the Galactic plane are clearly visible on the
right hand side of the image, while one should recall that
the principal Galactic crossing has been excised from the data.
It is clearly seen that there is apparent continuity of structure 
between adjacent independent 
data scans which are separated by less than the $8\dg$ beam
width (see the higher declination strips in the plot where the data are
more fully sampled). Where the data are not fully sampled (the lower
declinations) the MEM has reverted to zero as expected and this is
seen as `stripping' along declinations in the reconstructed map. 

In the next section we compare our reconstructed sky, reconvolved MEM scans
and stacked data scans with the expected CMB and foreground signals at this 
observing frequency. 

\section{Non-cosmological foreground contributions}
\label{noncos}

\subsection{Point sources}
\label{points}

\begin{figure}
\centerline{\hbox{\psfig{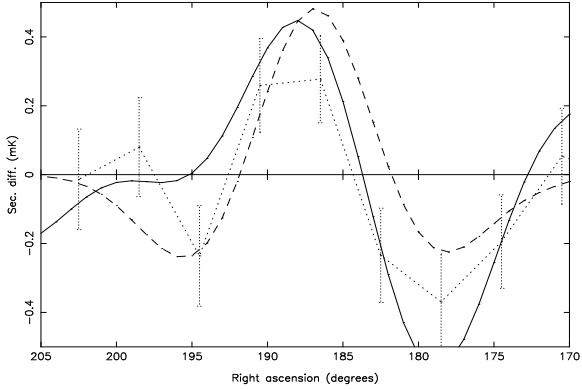}}}
\caption{Comparison between the MEM reconstructed sky convolved in the 
Tenerife beam (solid line), the predicted point source contribution at
Dec.=$1.1\dg$ (dashed line) and the Tenerife data (dotted line with one 
sigma error bars shown). The source observed is 3C273 
(RA$={12^h}{26^m}{33^s}$, Dec.$=+{02\dg}{19'}{43''}$).}
\label{fig:compar}
\end{figure}

We have estimated the contribution of discrete sources to the Tenerife 
10 GHz data using the
K\"{u}hr \et (1981) catalogue, the VLA calibrator list and the Green 
Bank sky surveys (Condon \& Broderick 1986); sources $\lta 1$ Jy at 10.4 GHz
were not included in the analysis. We have
modelled the response of our instrument to these point sources by
converting their fluxes into antenna temperature (1 Jy is equivalent to
12 $\mu$K for our experiment), convolving these with the triple beam of
our instrument and sampling as for the real data  (see the details
in Guti\'errez \et 1995). 
The two main radio sources at high Galactic latitude,
which we expect to see in the Tenerife scans are 
3C 273 (RA=$12^h26^m33^s$, Dec.=$+02\degg19^{\prime}43^{\prime\prime}$) 
with a flux density at 10
GHz $\sim 45$ Jy; this object should contribute with a peak amplitude
$\Delta T\sim 500$ $\mu$K in the triple beam to our data at Dec.=+1.1\degg, and 
3C84 (RA=$3^h16^m30^s$, Dec.=$+41\degg19^{\prime}52^{\prime\prime}$) with a
flux density at 10 GHz $\sim 51$Jy.
Figure \ref{fig:compar} presents a comparison between our MEM result reconvolved
in the Tenerife triple beam, the data and
the predicted contribution of the radio source 3C273. We will show below how
a diffuse Galactic contribution near the
position of this point source accounts for the differences in
amplitude and shape of the radio source prediction and our data.
The radio sources 3C 273 and 3C84 have also been detected in the deconvolved map
of the sky shown in Figure \ref{fig:8degpn}. For example, 3C 273 is seen with 
an amplitude of $1200 \pm 140\mu$K. Also clearly detected are 3C345 
(RA=$16^h41^m18^s$, Dec.=$+39\degg54^{\prime}11^{\prime\prime}$)and 4C39
(RA=$9^h23^m56^s$, Dec.=$+39\degg15^{\prime}23^{\prime\prime}$)
in both the reconvolved scans and the deconvolved map. Many other
sources are seen in the deconvolved map but these may be swamped by the 
Galactic emission so we cannot say with confidence that any 
originate from point sources. 
These could originate from the diffuse 
Galactic emission or from the CMB. For example, features at 
Dec.$\sim$+40\degg, RA$\sim $180\degg,
Dec.$\sim$+17.5\degg, RA$\sim $240\degg and Dec.$\sim$+1.1, RA$\sim 220\degg$ do 
not correspond to any known radio sources (see Figure \ref{fig:stack}). The additional
contribution by unresolved radio sources has been estimated to be
$\Delta T/T\sim 10^{-5}$ at 10.4 GHz (Franceschini \et 1989) in a single 
beam, and so will be less in the Tenerife switched beam, and is not
considered in the analysis presented here.

\subsection{Diffuse Galactic contamination}
\label{diffuse}

The contribution of the diffuse Galactic emission in our data can be
estimated in principle using the available maps at frequencies below
1.5 GHz. We have used the 408 MHz (Haslam \et 1982) and 1420 MHz (Reich
\& Reich 1988) surveys; unfortunately  the usefulness of these maps is
limited because a significant part of the high Galactic latitude
structure evident in them is due to systematic effects (Davies, Watson
\& Guti\'errez 1996b).  Only in regions (such as crossings of the
Galactic plane) where the signal dominates clearly over the
systematic uncertainties, is it possible to estimate the expected signals
at higher frequencies. With this in mind, we have converted these two
maps to a common resolution (1\degg x 1\degg in right ascension and
declination respectively) and convolved them with our triple beam
response.

We can compare this contribution at 408 and 1420 MHz, with our
data at 10.4 GHz to determine the spectral index of the Galactic
emission in the region where these signals are high enough to dominate
over the systematic effects in the low frequency surveys.  We assume
a power law spectra ($T \propto \nu^{-\beta}$) 
for the signal with an index independent of the
frequency, but varying spatially.  The signals in
the Galactic anti-centre are weaker than those for the Galactic plane
crossing and are mixed up with several extended
structures, but even in this case we can draw some conclusions about
the spectral index in this region; 
we obtain $\beta=3.0\pm 0.2$ between 408/1420 MHz and $\beta =
2.1\pm0.4$ between 1420/10400 MHz which indicates that free-free
emission dominates over synchrotron at frequencies $\gta 1420$ MHz.
One of the stronger structures in the region away from the galactic plane 
is in RA$\sim 12^h-13^h$, Dec$\sim 0$\degg~and
therefore the main contribution should be to our data at Dec=1.1\degg.
This structure at 408 MHz, assuming an spectral
index $\beta =2.8$, gives a predicted peak amplitude at 10.4 GHz of
$\sim 500$ $\mu$K; we believe that this is responsible for the
distortion between our measurements at Dec=1.1\degg and the predictions
for the radio source 3C 273. 

\section{Statistical analysis}
\label{stats}

We have analysed the statistical properties of the signals present in
our data using the likelihood function and a Bayesian analysis. 
This method has been widely
used in the past by our group (see {\it e.g.} Davies \et 1987) and
incorporates all the relevant parameters of the observations:
experimental configuration, sampling, correlation between measurements,
etc. The analysis assumes that both the noise and the signal follow a
Gaussian distribution  fully determined by their respective
auto-correlation function (ACF). The source of dominant noise in our
data is thermal noise in the receivers which is independent in each
data-point (Davies \et 1996a) and therefore it only contributes to
the terms in the diagonal of the ACF matrix. We have restricted our analysis
to data in which we have a minimum number of 10 independent
measurements for the full RA range (Dec. $7.3\degg$ does not have
enough data) and to data for which we have a point source prediction
(Dec. $-17.3\degg$ is not covered by the Green Bank survey). 
This region represents approximately 3000 square degrees on
the sky. Table~\ref{ta:tab2} presents the sensitivity per 
beam in the RA range used
in this analysis. Also column 4 gives the mean number of
independent measurements which contribute to each point. We emphasize
that this statistical analysis has been performed directly on the scan
data, and not on the MEM deconvolved sky map produced during the
baseline subtraction process. Thus, for this section, any effects of
using a MEM approach are restricted to the baselines subracted from
the raw data, which will not contain or affect any of the astronomical
information to which the likelihood analysis is sensitive. 

\begin{table}
\caption{Statistics of the data used in the analysis. 95\% confidence
limits are shown.}
\label{ta:tab2}
\centering
\begin{tabular}{ccccc} \hline
Dec. & RA  & $\sigma$ ($\mu$K)& Indep. &
 $(\frac{l(l+1)}{2\pi}C_l)^{1/2}\, (10^{-5}$) ($\mu$K) \\ 
+46.6\degg & 161\degg-250\degg & 116 & 15 & $\le 8.5$  \\
+42.6\degg & 161\degg-250\degg & 117 & 14 & $\le 10.3$\\
+39.4\degg & 176\degg-250\degg &  81 & 42 & $1.8^{+2.3}_{-2.0}$ \\
+37.2\degg & 161\degg-250\degg & 113 & 17 & $5.7^{+3.2}_{-2.9}$ \\
+27.2\degg & 161\degg-240\degg & 139 & 11 & $4.5^{+3.0}_{-3.9}$ \\
+17.5\degg & 171\degg-240\degg & 144 & 12 & $4.3^{+3.0}_{-4.4}$ \\
+1.1\degg &  171\degg-230\degg &  96 & 58 & $5.0^{+3.3}_{-3.0}$\\ \hline
\end{tabular}
\end{table}


We made two different analyses: the first considers the data of
each declination independently, and the 
second considers the full two-dimensional data set 
for the analysis. Due to the spherical nature of the sky we
expand the fluctuations using spherical harmonics $Y_l^m$ (see for
example Efstathiou 1989), 

\begin{equation}
{\Delta T\over T} (\theta,\phi)= \sum_{l,m} a_l^m Y_l^m (\theta,\phi).
\label{eq:expanse}
\end{equation}

\begin{figure}
\centerline{\hbox{\psfig{figure=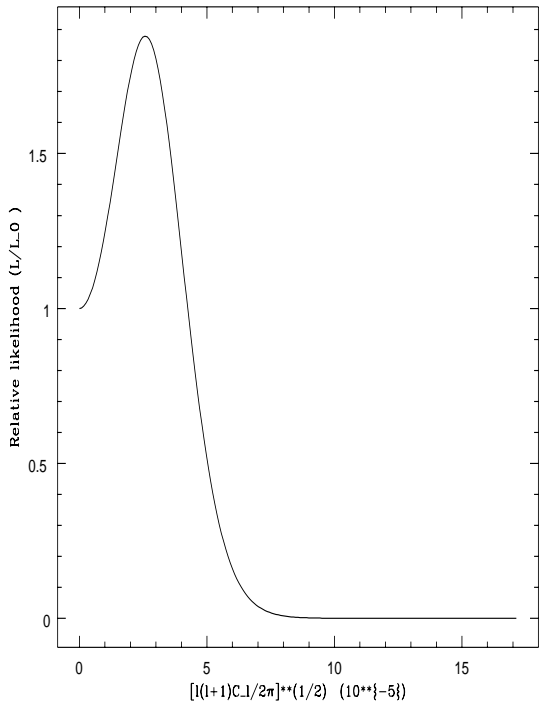,height=4.0in,width=3.0in,angle=0}}}
\caption{The likelihood function from the analysis of the full data set. There
is a clearly defined peak at 
$(\frac{l(l+1)}{2\pi}C_l)^{1/2}\, (10^{-5}) = {2.6^{+1.3}_{-1.6}}$
($95\% $ confidence level).
}
\label{fig:likel}
\end{figure}

\noindent 
A likelihood analysis was performed assuming a
Harrison-Zel'dovich spectrum, thus the parameter fitted for was the
cosmic quadrupole normalisation for the spherical harmonic expansion, 
$Q_{RMS-PS}$ (see Smoot \et 1992). 
Since the $l$ range sampled is small we can easily get an
equivalent flat band pass estimate for $C_l$ ($C_l = < |a_l^m|^2 >$)
and this is what is shown
in the Table \ref{ta:tab2}. The experiment has a peak sensitivity to an $l$ of
about 12. The fifth column of
Table~\ref{ta:tab2} gives, for the one-dimensional analysis, the amplitude of the
signal detected with the one-sigma confidence level. The confidence
limits on these signals were found by integration over a uniform prior
for the likelihood function. 
These analyses ignore correlations
between measurements at adjacent declinations. Therefore a full
likelihood analysis, taking this correlation into account, 
should constrain the signal more efficiently. It should be noted that 
the two dimensional analysis assumes that the signal has the same
origin over the full sky coverage but this may not be the case because
of the differing levels of Galactic signal between declinations and
across the RA range. In
Figure \ref{fig:likel} the likelihood function resulting from this
analysis is shown. It shows a clear, well defined peak at 
$(\frac{l(l+1)}{2\pi}C_l)^{1/2}\, (10^{-5}) = 2.6^{+1.3}_{-1.6}$ 
(95 \% confidence level). This value would
correspond to a value of $Q_{RMS-PS} = 45.2^{+23.8}_{-27.2}$ $\mu$K. 
Our results are compatible with the constraints on the signal
in each declination considered separately but it is clear that the two 
dimensional likelihood analysis improves the  constraints on the amplitude 
of the astronomical signal. 

We may compare the results obtained here with the amplitude of the CMB
structure found in Hancock \et (1994) at higher frequencies. They
found $Q_{RMS-PS}
\sim 21\mu$K in an $5\dg$ FWHM switched beam and taking into account
the extra dilution we expect a slightly lower level in an $8\dg$ FWHM
switched beam, assuming a $n=1$ power spectrum. We thus 
see that it is likely that the majority of the
signal in the 10 GHz, FWHM=$8\deg$ data is due to Galactic
sources. If we assume that the majority of the signal found here
is Galactic synchrotron or free--free 
and we use a spatial spectrum of $C_l \propto l^{-3}$ (estimated
from the Haslam \et 1982 maps) to predict the
expected galactic contamination in a $5\dg$ FWHM beam at 10 GHz, then using
a full likelihood analysis
we find that we expect an {\em rms} temperature across the scans of 
$\Delta T_{rms} = 55^{+32}_{-26}$ $\mu$K. We
note that this is an upper limit on the Galactic contribution to the
$5\dg$ data as the variability of the sources has been ignored when
the subtraction was performed (this results in a residual signal from
the point sources in the data during the likelihood analysis) and the
analysis also includes regions where the Galactic signal is expected
to be higher (for example the North Polar Spur). The $5\degg$ FWHM
Tenerife scans are centred on Dec. $40\degg$ and it can be seen from
Table \ref{ta:tab2} that this is the region with the lowest Galactic
contamination. The results 
reported in Gutierrez {\em et al} (1997), for the $5\degg$ FWHM, 10 GHz
Tenerife experiment, show that the signal found was $Q_{RMS-PS} <
33.8$ $\mu$K (corresponding to a signal of $\Delta T_{rms} < 53$
$\mu$K) which is consistent with our prediction (also taking into
account the more significant contribution from the CMB at $5\deg$). 
This comparison allows us to restrict the {\em maximum}
Galactic contribution to the signal found in Hancock \et 1994 to be
$\Delta T_{rms}  \sim 18 - 23\mu$K at 15 GHz and 
$\Delta T_{rms} \sim 2 - 4\mu$K at 33 GHz depending on whether the contamination is 
dominated by synchrotron or free-free emission. 

\section{Conclusions}

We have presented here a new method for analysing the data from microwave
background experiments. 
As seen from simulations performed in Section \ref{sim}, the
positive/negative MEM algorithm performs very well recovering the amplitude,
position and morphology of structures 
in both the reconvolved scans and the two-dimensional deconvolved sky
map. We conclude that no bias, other than the `damping' enforcement,
is introduced into the results from the methods described here, as 
the bare minimum of prior knowledge of the sky is
required. A simultaneous baseline fit is also possible.
Even with the lowest signal to noise ratio (the $100\mu$K
noise simulation which corresponds to our worse case in 
the Tenerife experiments) all of the main features on the sky were 
reconstructed. Using this method we are able to put constraints on the
galactic contamination for other experiments at higher frequencies which is
essential when trying to determine the level of CMB fluctuations present.  

It is clear that this approach works well and provides a useful
technique for extracting the optimum CMB maps from both current and
future multi-frequency experiments. This will become of ever
increasing importance as the quality of CMB experiments improves. 
At present we are using this method to analyse the new data from the three
beam switching Tenerife experiments at 10 GHz, 15 GHz and 33 GHz with
angular resolutions of $\sim 5\degg$. We hope the two dimensional 
results from these will be available shortly. 

\subsection*{ACKNOWLEDGEMENTS}

\noindent The Tenerife experiments are supported by the UK Particle
Physics and Astronomy Research Council, the European Community Science
programme contract SCI-ST920830, the Human Capital and Mobility
contract CHRXCT920079 and the Spanish DGICYT science programme. A.W.
Jones wishes to acknowledge a UK Particle Physics
and Astronomy Research Council Studentship. S.
Hancock wishes to acknowledge a Research Fellowship at St. John's
College, Cambridge, UK. G. Rocha wishes to acknowledge a JNICT Studentship
from Portugal. 

{}

\newpage

{\Large \bf Figure captions}

Figure 1: The 15 scans obtained at Dec $=46.6\dg $ 
displayed as a function of right ascension. Each plot shows 
the second difference in mK after binning into $1\dg $ 
bins. A running mean has been subtracted from each scan. 
Long scans are displayed modulo $360\dg $.

Figure 2: The data from scan 5 of Figure \ref{fig:typical15}
displayed on an expanded temperature scale against RA bin number. Long 
timescale variations in the mean level are evident in the RAW scan
(bottom panel). The middle panel shows the 
baseline fit found by the method of Section \ref{memalg}.
The top panel shows the
baseline corrected scan. The bin numbers exceed 360 since 
the scan begins near the end of an LST day, and the data are not folded modulo 
$360\dg $.

Figure 3: The transfer function for the Tenerife experiments.

Figure 4: The  solid line shows the sky simulation convolved
with the Tenerife
$8\dg$ experiment. The bold dotted line in the top figure 
shows the MEM reconstructed sky after
reconvolution with the Tenerife beam, averaged over 
simulations of the  MEM output from a simulated experiment with $25\mu$K
Gaussian noise added to each scan. Also shown are the $68\%$ confidence limits 
(dotted lines) on this
reconvolution. The bold dotted line in the bottom figure 
shows the reconvolution averaged over 
simulations with $100\mu$K Gaussian noise added to each scan. 
The $68\%$ confidence
bounds (dotted lines) are also shown for this scan.

Figure 5: The top figure is the simulated sky convolved with an $8\degg$
Gaussian beam. The middle and bottom figures are the reconstructed skies from
that simulation after scans with $25\mu$K and $100\mu$K noise levels
respectively, simulating the Tenerife experiment, were made and passed through
MEM. They are all convolved with a final Gaussian of the same size.

Figure 6: Histograms of the errors in the reconstruction of the
simulated sky maps. (a) shows the 
${{\rm T}_{recon}-{{\rm T}_{input}}}\over {{\rm T}_{input}}$ for the
$25\mu$K noise simulation and (c) shows the integrated 
$\left| {{{\rm T}_{recon}-{{\rm T}_{input}}}\over {{\rm T}_{input}}} \right|$ 
for (a). (b) and (d) are the corresponding plots for the
$100\mu$K noise simulation. 

Figure 7: The stacked scans at each declination displayed as a function 
of right ascension. Again the plots are the second difference binned 
into $4\dg$ bins and the $68\%$ confidence limits. The main Galactic
plane crossing has been excluded, and only positions on the sky in
which we have more than $\sim$10 independent measurements have been
plotted. Also shown (solid line) is the reconvolved result from MEM overlayed 
onto each declination scan.

Figure 8: MEM reconstruction of the sky at 10.4 GHz, as seen by the Tenerife
$8.4\dg $ FWHM experiment.

Figure 9: Comparison between the MEM reconstructed sky convolved in the 
Tenerife beam (solid line), the predicted point source contribution at
Dec.=$1.1\dg$ (dashed line) and the Tenerife data (dotted line with one 
sigma error bars shown). The source observed is 3C273 
(RA$={12^h}{26^m}{33^s}$, Dec.$=+{02\dg}{19'}{43''}$).

Figure 10: The likelihood function from the analysis of the full data set. There
is a clearly defined peak at 
$(\frac{l(l+1)}{2\pi}C_l)^{1/2}\, (10^{-5}) = {2.6^{+1.3}_{-1.6}}$
($95\% $ confidence level).

\end{document}